\documentclass[aps,pre,showpacs, twocolumn]{revtex4-1}
\usepackage[utf8]{inputenc}
\usepackage{graphicx}
\usepackage[caption=false]{subfig}
\graphicspath{{./images/}}
\usepackage{textcomp,amsmath,amssymb}
\usepackage{hyperref}
\hypersetup{pdftitle={Saturation of number variance in embedded random matrix ensembles},
           pdfauthor={Ravi Prakash, Akhilesh Pandey},
           pdfcreator={pdflatex from texlive-2013},
           colorlinks=true,
           citecolor=blue,
           urlcolor=blue}

\newcommand{\Si}{\mbox{Si}}
\newcommand{\Ci}{\mbox{Ci}}
\newcommand{\pFq}[5]{_#1F_#2 \left[ 
\begin{array}{c}
#3 \vspace{1mm}\\
#4
\end{array} ;#5 
\right]}

\begin{document}
\title{Saturation of number variance in embedded random matrix ensembles}
\author{Ravi Prakash} \email{raviprakash.sps@gmail.com}
\author{Akhilesh Pandey}\email{ap0700@mail.jnu.ac.in}\email{apandey2006@gmail.com}
\affiliation{School of Physical Sciences, Jawaharlal Nehru University, New Delhi -- 110067, India}

\begin{abstract}
We study fluctuation properties of embedded random matrix ensembles of non-interacting particles. For ensemble of two non-interacting particle systems, we find that unlike the spectra of classical random matrices, correlation functions are non-stationary. In the locally stationary region of spectra, we study the number variance and the spacing distributions. The spacing distributions follow the Poisson statistics which is a key behavior of uncorrelated spectra. The number variance varies linearly as in the Poisson case for short correlation lengths but a kind of regularization occurs for large correlation lengths, and the number variance approaches saturation values. These results are known in the study of integrable systems but are being demonstrated for the first time in random matrix theory. We conjecture that the interacting particle cases, which exhibit the characteristics of classical random matrices for short correlation lengths, will also show saturation effects for large correlation lengths.
\end{abstract}


\maketitle
\section{Introduction}
Random matrix theory has become a great tool to study energy level statistics of quantum chaotic systems,  such as complex nuclei, mesoscopic transport,  Sinai billiard, etc \cite{nuclear, rmp, mlmehta, bg, rmp-beenakker, efetov1999, phyrep-guhr, stockmann, haakebook, forrester-book, oxford}. It is also applicable to time series problems \cite{timeseries1, timeseries2, timeseries3}, communication theory \cite{kp, wireless} etc. The local level fluctuations are universal and depend only on the symmetry class of the system. It is conjectured and confirmed by numerical and experimental results that quantum chaotic systems follow Wigner repulsion whereas integrable systems exhibit Poisson statistics \cite{berrytabor, hpb, bgs, berry}.

Although random matrix theories successfully describe many universal properties, they are valid only for short correlation lengths with respect to the size of the spectra. The number variance statistic, $\Sigma^2(r)$, exhibits logarithmic and linear behavior at such correlation lengths for quantum chaotic and integrable systems respectively; here $\Sigma^2(r)$ is the variance of the number of levels in intervals of length $rD$, where $D$ is the local average spacing.

A numerical analysis of the spectrum of quantum rectangular billiard shows that while $\Sigma^2(r)$ starts linearly for small $r$ as in the Poisson case, it eventually becomes a constant \cite{casati2}. Berry \cite{berry} has shown by semi-classical methods that this is a generic behavior of quantum integrable systems. He has also predicted that a similar saturation effect will be seen in the spectra of quantum chaotic systems. Thus the saturation phenomenon is universal for real quantum systems, but the saturation value varies from system to system. We also mention that the saturation phenomenon has been found for the Riemann-$\zeta$ function \cite{saturation-2, saturation-3}.

It is believed that the saturation phenomenon will not be seen in random matrix ensembles. This is certainly true for the Gaussian, circular and other related invariant ensembles. Our purpose in this article is to show that the embedded ensembles (EE) \cite{FW-1970, embedded, embedded-3, embedded-2, rmp, kota} exhibit saturation effects. Embedded ensembles are ensembles of $m$-particle systems with interactions of rank $k$ (i.e., $k$-particle interactions), where $k < m$. These are applicable to the study of statistical properties of many particle systems. We mention that the saturation phenomenon in embedded ensemble has been predicted \cite{pandey1986}, but has never been verified. This conjecture was based on the observation that very short wavelength fluctuations in the embedded ensembles are suppressed \cite{embedded-2, rmp-note}.

In this article, we consider the embedded random matrix ensemble of two non-interacting particles, i.e., $k = 1$ and $m = 2$. We derive expressions for the spectral density and the two-level correlation function and show that the level fluctuation statistics are non-stationary. We also confirm that the very short wavelength fluctuations are indeed suppressed. We study the number variance statistics which display linear behavior for short correlation lengths, as it should for integrable systems, but approaches saturation as we go towards higher lengths, confirming thereby the above mentioned conjecture. We believe that the saturation will also be found in the interacting case, but this is difficult to demonstrate.

This article is organized as follows.  In section \ref{egxe-sec-EGE}, we describe the EEs of non-interacting particles and their $n$-level correlation functions. We derive expressions for the spectral density and the two-level correlation function in sections \ref{egxe-sec-R1} and \ref{egxe-sec-correlation} respectively. These expressions are required for the unfolding of the spectrum and the evaluation of the number variance as well as the spacing distribution. In section \ref{egxe-sec-formfactor}, we study the spectral form factor and the number variance. The numerical method to study the form factor and the number variance is discussed in section \ref{egxe2-sec-mc}. The short-range behavior is studied in terms of nearest neighbor spacing distributions as well as number variance and is discussed in section \ref{egxe2-sec-sapacings}. The saturation phenomenon at large correlation lengths is discussed in section \ref{egxe-sec-sigmasq-sat}. In sections \ref{egxe2-sec-sigmasq-EGUE}, \ref{egxe2-sec-sigmasq-EGSE}, \ref{egxe2-sec-sigmasq-EGOE}, we obtain the number variance and their saturation values for the unitary, symplectic and orthogonal cases of EEs respectively. In these sections we have taken $1000$-dimensional one-particle spectra with constant average density. To improve the statistics significantly we consider $5000$-dimensional one-particle spectra in section \ref{egxe-sec-sigmasq-N5000}. In section \ref{egxe2-sec-sigmasq-nonuniform}, we consider an example of non-uniform average density. We summarize our results in section \ref{egxe-sec-summary}. In Appendices \ref{appendix-rn-2p}-\ref{appendix-sigmasq-egoe}, some details of the analytical calculations are given. 

\section{Embedded Random Matrix Ensembles} \label{egxe-sec-EGE}
For a system of $m$ non-interacting fermions, the energy levels can be defined as,
\begin{align}
\label{egxe-e-2p}
E = \sum n_j \epsilon_j,
\end{align}
where the $\epsilon_j$ are the single particle energy levels and the $n_j$ are occupancy numbers with values $0,1$ such that $\sum n_j = m$. The boson case can also be considered similarly. The single particle Hamiltonian is modeled by Gaussian ensemble. We consider all three types viz., Gaussian orthogonal (GOE), Gaussian unitary (GUE) and Gaussian symplectic (GSE) ensembles. We shall refer to the corresponding embedded ensembles as non-interacting embedded GOE (EGOE), embedded GUE (EGUE) and embedded GSE (EGSE) respectively \cite{rmp, embedded-2}. There are $N$ single particle energy levels. The dimension of the $m$-particle spectra will be,
\begin{equation}
d_m = \frac{N!}{m! (N-m)!}.
\end{equation}
Thus for two particles, it will be $d_2 = N(N-1)/2$.

Let $P(x_1,\ldots,x_N)$ be the joint probability distribution (jpd) of the single-particle levels. Then the one-particle $n$-level correlation function, $R_n(x_1, \ldots, x_n)$ is defined by
\begin{equation}
R_n = \frac{N!}{(N-n)!} \int\ldots\int P(x_1,\ldots,x_N) ~ \mathrm dx_{n+1} \ldots \mathrm dx_N.
\end{equation}
The $m$-particle correlation functions have similar definitions with $N$ replaced by $d_m$.

We write correlation functions in terms of cluster functions. The $n$-level correlations, $R_n$ can be written in terms of cluster function $T_n$ as follows\cite{mlmehta, dyson3}
\begin{equation}
\label{egxe-rn-tn}
R_n = \sum\limits_G (-1)^{n-q} \prod\limits_{j=1}^{q}T_{G_j}(x_k, \mbox{with } k \mbox{ in } G_j),
\end{equation}
where $G$ stands for any division of the indices $(1,\ldots,n)$ into $q$ subgroups $(G_1,\ldots,G_q)$. For example, the one and two-level correlation functions can be written as,
\begin{equation}
\label{egxe-r1-t1}
R_1(x) = T_1(x)
\end{equation}
and
\begin{equation}
\label{egxe-r2-t2}
R_2(x_1, x_2) = T_1(x_1) T_1(x_2) - T_2(x_1, x_2).
\end{equation}

Since we are dealing with non-interacting particles, the $n$-level correlation function for the two-particle spectra can be written in terms of one-particle $p$-level correlation functions, where $p\leq 2n$. We denote the one-particle and two-particle correlation functions by $R_n$ and $R_n^{(2)}$ respectively. A similar convention will also be used for other physical quantities unless or otherwise stated explicitly.

We write correlation functions for single particle spectrum in terms of cluster functions. The detailed derivation is provided in Appendix \ref{appendix-rn-2p}. We get following expressions for spectral density and two-level correlation function for two-particle spectrum,
\begin{align}
\label{egxe-R1-2p-t2}
\nonumber
R_1^{(2)}(\xi) & = \frac{1}{2}  \int_{-\infty}^{\infty}   \int_{-\infty}^{\infty} \left[T_1(x_1)T_1(x_2) - T_2(x_1, x_2)\right] \\
&\times \delta\left(\xi - \left(x_1+x_2\right)\right) ~\mathrm dx_1 \mathrm dx_2.
\end{align}
The two-level correlation function is given by,
\begin{align}
\label{egxe-R2-2p-r4}
\nonumber
R_2^{(2)}\left(\xi_1,\xi_2\right) & = \int_{-\infty}^{\infty}  \int_{-\infty}^{\infty} \int_{-\infty}^{\infty} \int_{-\infty}^{\infty}  \bigg[\frac{1}{4} R_4(x_1,x_2,x_3,x_4) \\
\nonumber
& + R_3(x_1,x_2,x_3) \delta(x_1 - x_4) \bigg]
\delta\big(\xi_1 - (x_1 + x_2)\big) \\
& \times \delta\big(\xi_2 - (x_3 + x_4)\big) ~ \mathrm{d}x_1 \mathrm{d}x_2 \mathrm{d}x_3 \mathrm{d}x_4.
\end{align}
$R_2$ and $R_1$ terms do not appear in (\ref{egxe-R2-2p-r4}) because we are dealing with fermions. The expression can also be written in terms of cluster functions by substituting (\ref{egxe-r3-t3}) and (\ref{egxe-r4-t4}) in (\ref{egxe-R2-2p-r4}).

\section{Spectral Density for the Two-Particle Spectrum} \label{egxe-sec-R1}
\begin{figure*}
\includegraphics[width=\linewidth, clip]{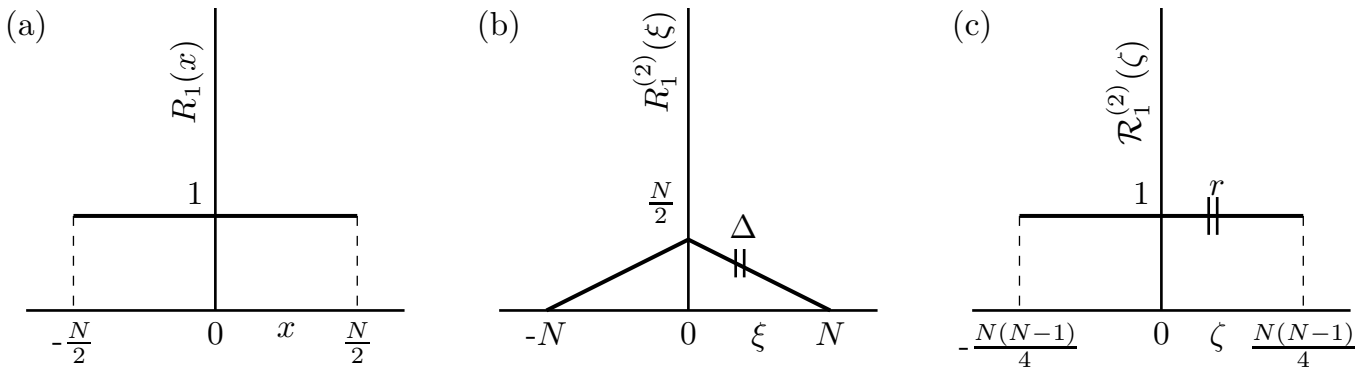}
\caption{Schematic representation of (a) Spectral density for an unfolded one-particle spectrum, (b) Spectral density for corresponding two-particle spectrum and (c) Spectral density for the unfolded two-particle spectrum. \label{egxe-fig-density-all}}
\end{figure*}

 We consider single particle spectrum as unfolded spectrum so that the average spectral density is uniform (Fig.\ref{egxe-fig-density-all}a). The non-uniform case is briefly discussed in the section \ref{egxe2-sec-sigmasq-nonuniform}. The average density, $R_1$, is taken as,
\begin{align}
\label{egxe-r1-unfolded}
R_1(x) & = \left\{ \begin{array}{ll} 
1, & -\frac{N}{2} \leq x \leq \frac{N}{2},  \\
0, & \mbox{otherwise.}
\end{array}
\right.
\end{align}
For the Gaussian ensembles mentioned above, the correlation and the cluster functions are translationally invariant, and therefore, $T_2$ is a function of the difference of its arguments \cite{apandey}. We, therefore, introduce new cluster function $Y_2$ for the unfolded spectrum,
\begin{equation}
\label{egxe-y2}
Y_2(x_1- x_2) = T_2(x_1,x_2).
\end{equation}
By using (\ref{egxe-y2}), the two-particle spectral density given in (\ref{egxe-R1-2p-t2}) can be written as,
\begin{align}
\label{egxe-R1-2p-y2}
\nonumber
R_1^{(2)}(\xi) & = \frac{1}{2}  \int_{-N/2}^{N/2}   \int_{-N/2}^{N/2} \big[T_1(x_1)T_1(x_2) - Y_2(x_1-x_2)\big] \\ & \times \delta\big(\xi - (x_1+x_2)\big) ~\mathrm dx_1 \mathrm dx_2.
\end{align}
Integral of the first term is the convolution of $T_2$ with itself. Integration of the second term depends on the ensemble. However in the large $N$ limit and away from the end of the spectrum $(|\xi| <N)$, this integral is equal to half of the normalization of $Y_2$. Substituting (\ref{egxe-r1-unfolded}) in (\ref{egxe-R1-2p-y2}), we get,
\begin{equation}
\label{egxe-R1-2p}
R_1^{(2)}(\xi)=  \left\{ \begin{array}{ll}
\frac{N}{2}-\frac{|\xi|}{2} - \frac{1}{4}, & -N \leq \xi \leq N, \\
0, & \mbox{otherwise.} \end{array}\right.
\end{equation}
Above expression describes the spectral density for two particle system (Fig.\ref{egxe-fig-density-all}b). For large $N$, the last term (i.e., $-1/4$) can be dropped. We will use this approximation to unfold the two-particle spectrum.

\section{Two-Level Correlation function for the Two-Particle Spectrum} \label{egxe-sec-correlation}
The two-level correlation function for the two-particle spectrum given in (\ref{egxe-R2-2p-r4}) can be simplified considerably for large $N$. See Appendix \ref{appendix-r2-2p} for detailed calculation. We find that the two-level cluster function is given by, 
\begin{equation}
\label{egxe-T2-2p}
T_2^{(2)}(\xi_1,\xi_2) = \frac{N - |\xi|}{2}\left[2 Y_2(\Delta) - \int_{-\infty}^{\infty}Y_2(x)Y_2(\Delta - x) \mathrm dx\right],
\end{equation}
where
\begin{equation}
\label{egxe-xi-def}
\xi = \frac{\xi_1 + \xi_2}{2},~\Delta = \xi_2 - \xi_1
\end{equation}
See Fig. \ref{egxe-fig-density-all} for $\xi$ and $\Delta$. We consider $|\Delta| = O(1)$, which means that the number of two-particle energy levels in $|\Delta|$ is $O(N)$.

To study the local fluctuations we unfold the two-particle spectrum by introducing the new variable $\zeta$ defined by,
\begin{equation}
\label{egxe-zeta-xi}
\zeta = \int_0^{\xi} R_1^{(2)}(\xi^\prime) ~\mathrm d\xi^\prime.
\end{equation}
The average spectral density, $\mathcal R_1^{(2)}(\zeta)$, in the new variable, $\zeta$,  is one (Fig.\ref{egxe-fig-density-all}c).
The average number of level in the interval $[\xi_1, \xi_2]$ is $|r|$, where $r$ is given by,
\begin{align}
\label{egxe-zeta-xi-diff}
r = \zeta_2 - \zeta_1  = R_1^{(2)}(\xi) \Delta.
\end{align}
The unfolded two-level cluster function is defined as,
\begin{align}
\label{egxe-Y2-2p-T2-2p}
Y_2^{(2)}(r;\xi)  \equiv \frac{T_2^{(2)}(\xi_1,\xi_2)}{R_1^{(2)}(\xi_1) R_1^{(2)}(\xi_2)}.
\end{align}
$Y_2^{(2)}(r;\xi)$ is given in terms of the one-particle cluster function as,
\begin{equation}
\label{egxe-Y2-2p}
Y_2^{(2)}(r;\xi) = \frac{1}{R_1^{(2)}(\xi)}\left[2 Y_2(\Delta) - \int_{-\infty}^{\infty}Y_2(x)Y_2(\Delta - x) ~\mathrm dx\right],
\end{equation}
where $\Delta$ should be written in terms of $r$ as in (\ref{egxe-zeta-xi-diff}). The density in a localized region can be assumed to be constant and therefore $R_1^{(2)}(\xi_1) \approx R_1^{(2)}(\xi_2) \approx R_1^{(2)}(\xi)$. The cluster function, $Y_2^{(2)}(r;\xi)$ is non-stationary globally as it depends on $\xi$. 
Note that (\ref{egxe-Y2-2p}) holds only for uniform single-particle density. Our results given in sections \ref{egxe-sec-formfactor}-\ref{egxe-sec-sigmasq-N5000} are all for the uniform case. We consider the non-uniform case briefly in section \ref{egxe2-sec-sigmasq-nonuniform}.

\section{Spectral Form Factor and Number variance} \label{egxe-sec-formfactor}
The two-level form factor $b_2^{(2)}(k;\xi)$ (or the Fourier transform of $Y_2^{(2)}(r;\xi)$) \cite{mlmehta, formfactor} for the two-particle system,
\begin{equation}
\label{egxe-B2-def}
b_2^{(2)}(k;\xi) = \int_{-\infty}^{\infty} Y_2^{(2)}(s;\xi) ~e^{2\pi i k s} ~\mathrm ds,
\end{equation}
is of special interest, as it will help us express number variance in a compact form. It can be written in terms of one-particle form factor (i.e., the Fourier transform of $Y_2(s)$),
\begin{equation}
b_2(k) = \int_{-\infty}^\infty Y_2(s) e^{2 \pi i k s} \mathrm ds.
\end{equation}
The Fourier transform of the first term in (\ref{egxe-Y2-2p}) gives $2b_2\left(k R_1^{(2)}(\xi)\right)$. The second term is the convolution of $Y_2$ with itself. Therefore its Fourier transform yields $\left[b_2\left(k R_1^{(2)}(\xi)\right)^2\right]$. Thus spectral form factor for the two-particle spectrum can be written as,
\begin{equation}
\label{egxe-B2-2p}
b_2^{(2)}(k;\xi) = 2 b_2\left(kR_1^{(2)}(\xi)\right) - b_2\left(kR_1^{(2)}(\xi)\right)^2.
\end{equation}
Eqs.(\ref{egxe-Y2-2p}) and (\ref{egxe-B2-2p}) connect the cluster function with one-particle cluster function.

It will be convenient to introduce the two-point form factor $F_2^{(2)}$ as,
\begin{equation}
F_2^{(2)}(k;\xi) = 1- b_2^{(2)}(k;\xi)
\end{equation}
which is the Fourier transform of the two-point correlation function,  
$S_2^{(2)} (r;\xi) = \delta(r) - Y_2^{(2)}(r;\xi)$,
with $\delta(r)$ being the self-correlation term.
Substituting in (\ref{egxe-B2-2p}), we get,
\begin{align}
\label{egxe-S2-2p}
\nonumber
F_2^{(2)}(k;\xi) & = \left[ 1- b_2\left( k R_1^{(2)}(\xi)\right) \right]^2 \\
& = \left[F_2\left(kR_1^{(2)}(\xi)\right)\right]^2,
\end{align}
where $F_2(k) = 1-b_2(k)$ represents the Fourier transform of the two-point correlation function, $S(r)$, for the corresponding one-particle spectrum.

The variance $\Sigma^2(r;\xi)$ of the number of levels, $n$, in the interval $[\zeta_1, \zeta_2]$ with $\zeta_2 > \zeta_1$ is defined as,
\begin{equation}
\label{egxe2-sigmasq-def}
\Sigma^2(r;\xi) = \overline{n^2} - \overline{n}^2,
\end{equation}
where $r = \overline{n}$. The bar denotes the ensemble average. The number variance can be calculated from the two-level cluster function and is given by \cite{mlmehta, rmp}, 
\begin{equation}
\label{egxe2-sigmasq-Y2}
\Sigma^2(r;\xi) = r -2 \int_0^r (r - s)Y_2^{(2)}(s;\xi)~\mathrm ds.
\end{equation}
It can also be written in terms of the two-point form factor as \cite{dyson6},
\begin{align}
\label{egxe2-sigmasq-B2}
\Sigma^2(r;\xi)
 = \int_{-\infty}^{\infty} F_2^{(2)}(k;\xi)\left(\frac{\sin(\pi k r)}{\pi k}\right)^2 ~\mathrm dk.
\end{align}
We will use the above expression for evaluating the number variance for the EEs.

\section{Numerical Study} \label{egxe2-sec-mc}
We construct the spectra of two non-interacting particle system numerically by using unfolded Gaussian ensemble spectra in (\ref{egxe-e-2p}). The one-particle spectra lie in the interval $[-N/2, N/2]$ with uniform average density. The spectral density for the two-particle system is a tent  like function in the interval $[-N, N]$ as shown in Fig.\ref{egxe-fig-density-all}b. We then unfold the two-particle spectra using the unfolding relation given in (\ref{egxe-zeta-xi}) so that the average spectral density is equal to $1$ (Fig.\ref{egxe-fig-density-all}c). Since spectral correlations are non-stationary, we consider a part of the tent spectrum which is small enough so  that the density does not vary significantly and also large enough to have approximately $O(N)$ levels. We verify in the later sections the two-particle form factor $F_2^{(2)}(k;\xi)$ and number variance for all three embedded ensembles discussed in Section \ref{egxe-sec-EGE}.

To calculate the form factor, $F_2^{(2)}(k;\xi)$ function, we consider the level correlation in the Fourier space. The two-level form factor can be written as,
\begin{equation}
\label{egxe2-S2}
F_2^{(2)}(k;\xi) = \frac{1}{N} \left[ \overline{\left| \sum_j e^{2\pi i k \zeta_j}\right|^2} -  \left | \overline{\sum_j e^{2\pi i k \zeta_j}} \right|^2 \right],
\end{equation}
where the $\zeta_j$ are levels of the unfolded spectra. We numerically evaluate the right-hand side for spectra under consideration. The expression can be verified analytically by evaluating the sum on the right-hand side in the continuum limit.
The number variance is calculated by using (\ref{egxe2-sigmasq-def}).

In this article, our numerical samples consist of ensembles having $100000$ spectra of the two-particle system. The spectra are constructed from all three types of Gaussian ensembles, each of dimension $N = 1000$. We have done calculations for three different regions of the spectra denoted by $\xi = 3.0, 105.6$ and $225.5$. These values correspond to $\zeta = 1500, 50000$ and $100000$ respectively in the corresponding unfolded spectra. We have also considered matrices with $N = 5000$ but with a smaller ensemble having $5000$ spectra in each case and $\xi = 20.0$ (with $\zeta = 50000$).

\section{Short range behavior} \label{egxe2-sec-sapacings}
\begin{figure}
\includegraphics[width=\linewidth, clip]{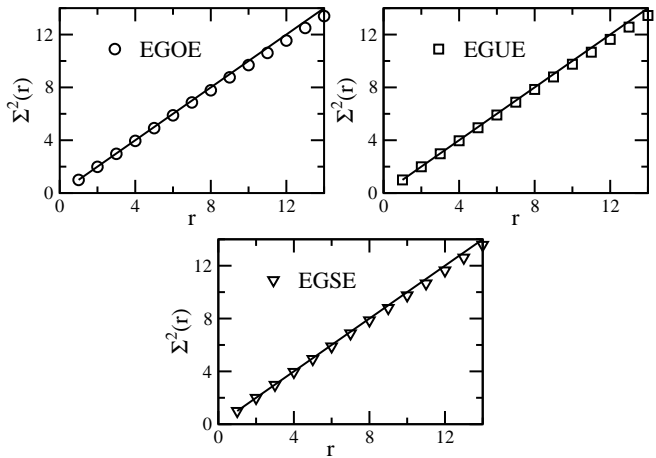}
\caption{Number variance at short correlation lengths for all three types of EEs. Agreement with Poisson result, shown by solid lines, is excellent.\label{egxe-fig-sigmasq-smr-all}}
\end{figure}

The two-particle spectrum is constructed by addition of energy levels of one particle spectrum. The eigenvalues exhibit the characteristics of uncorrelated spectrum. Therefore spectra of non-interacting particles should display the Poisson statistics for small correlation lengths -  a characteristic of  integrable systems.
We first consider the number variance to illustrate the short range behavior. From (\ref{egxe-S2-2p}, \ref{egxe2-sigmasq-B2}) one can predict that for short correlation lengths, viz. $r \ll N$, Poisson behavior will be obtained. The one-particle form factor $F_2(k)$ for the Gaussian ensembles increases from zero and goes to one for $|k| \gtrsim 1$. Eq. (\ref{egxe-S2-2p}) implies that for two-particle spectra $F_2^{(2)}(k;\xi)$ is one for  $k R_1^{(2)} \gtrsim 1$ or $k \gtrsim 1/N$ because the spectral density for the two-particle spectra is of the order $N$. See Fig.\ref{egxe2-fig-b2-EGUE}, Fig.\ref{egxe2-fig-b2-EGSE} and Fig.\ref{egxe2-fig-b2-EGOE} ahead. By substituting, $F_2^{(2)}(k;\xi) = 1$ in (\ref{egxe2-sigmasq-B2}), we obtain the Poisson result in the large $N$ limit,
\begin{align}
\label{egxe-sigmasq-sm-r}
\nonumber
\Sigma^2(r;\xi) & = 2 \int_0^{\infty} \left(\frac{\sin(\pi k r)}{\pi k}\right)^2 ~\mathrm dk \\
& = r.
\end{align}
This is shown in Fig.\ref{egxe-fig-sigmasq-smr-all} for all three types of EEs.

\begin{figure*}[!ht]
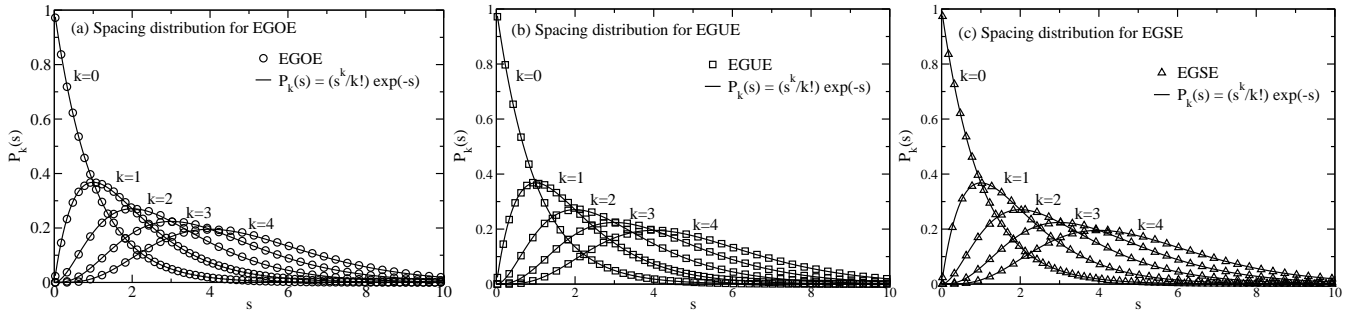

\subfloat{\includegraphics[width= 0.33\textwidth, clip]{spacing-egoe.eps}} 
\subfloat{\includegraphics[width= 0.33\textwidth, clip]{spacing-egue.eps}} %\\ [0ex] 
\subfloat{\includegraphics[width= 0.33\textwidth, clip]{spacing-egse.eps}}
\caption{$k^\text{th}$-nearest neighbor spacing distribution of two particle unfolded spectra. $k = 0$ denotes the nearest neighbor spacing distribution. Solid lines show the theoretical results for the normalized Poisson distribution. \label{egxe-fig-sp}}
\end{figure*}

Next we consider the nearest neighbor spacing distribution. The $k^\text{th}$-nearest neighbor spacing distribution $P_k(s)$ viz. the distribution of spacing between two levels with $k$ levels in the middle, exhibits Poisson statistics. $P_k(s)$ for integrable systems is given by,
\begin{equation}
P_k(s) = \frac{s^k}{k!} e^{-s}.
\end{equation}
Our results are shown in Fig.\ref{egxe-fig-sp} for all three types of EEs, viz., EGOE, EGUE and EGSE.

The agreement with the Poisson result is excellent for the number variance as well as the spacing distribution.

\section{Saturation phenomenon in number variance} \label{egxe-sec-sigmasq-sat}
\begin{figure}
\includegraphics[width=\linewidth, clip]{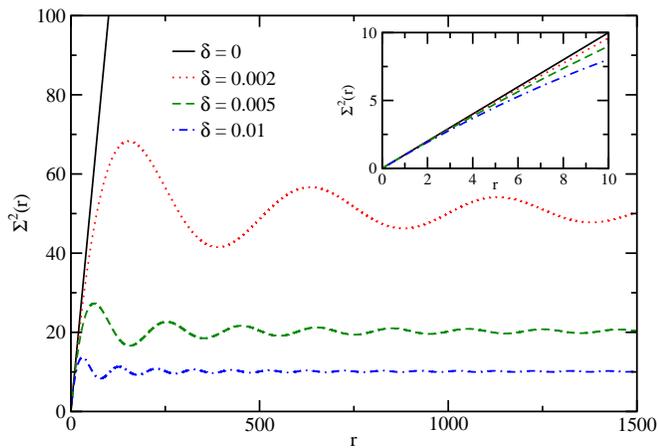}
\caption{(Color online) The number variance for the form factor given in (\ref{egxe-f2-model}) for several values of $\delta$. Inset shows the same figure for small correlation lengths. \label{egxe-fig-sigmasq-poisson}}
\end{figure}

Saturation of number variance for large correlation lengths, $r \gtrsim N$, comes from the suppression of fluctuations of short-wave frequencies (or large wavelengths). To explain it, we consider a model where the fluctuations of frequencies $< \delta$ are completely absent. The two-point form factor is given by
\begin{align}
\label{egxe-f2-model}
\mathcal F_2(k) = \left\{
\begin{array}{l l}
0, & 0 \leq |k| \leq \delta, \\
1, & \delta < |k| < \infty.
\end{array}
\right.
\end{align}
The number variance is evaluated using (\ref{egxe2-sigmasq-B2}) for the above form factor and is shown in Fig.\ref{egxe-fig-sigmasq-poisson} for several values of $\delta$. For $\delta = 0$, Poisson result is obtained. For $\delta > 0$, the number variance is Poisson-like for small $r$ and approaches saturation value $1/(\pi^2 \delta)$ for large $r$; see (\ref{egxe-sigmasq-model-gxe}) ahead.

For saturation to occur in the  general case, (\ref{egxe2-sigmasq-B2}) should converge for large values of $r$, i.e., $F_2^{(2)}(k;\xi) = O(|k|^\epsilon)$ for small $|k|$ with $\epsilon > 1$. The saturation value can now be evaluated by replacing $(\sin(\pi k r))^2$ in (\ref{egxe2-sigmasq-B2}) by $1/2$, its average over $r$ for large $r$. Thus, we have,
\begin{equation}
\label{egxe2-sigmasq-sat}
\Sigma^2_{\mathrm{sat}}(r;\xi) = \int_{-\infty}^\infty F_2^{(2)}(k;\xi) \frac{1}{2\pi^2 k^2} ~\mathrm d k.
\end{equation}
This will be verified ahead for systems of two non-interacting particles in sections \ref{egxe2-sec-sigmasq-EGUE}, \ref{egxe2-sec-sigmasq-EGSE}, \ref{egxe2-sec-sigmasq-EGOE} for EGUE, EGOE and EGSE respectively by rigorous calculations. Note that the saturation does not apply to the Gaussian and Poisson ensembles because in these cases (\ref{egxe2-sigmasq-sat}) is not convergent.

\begin{figure*}
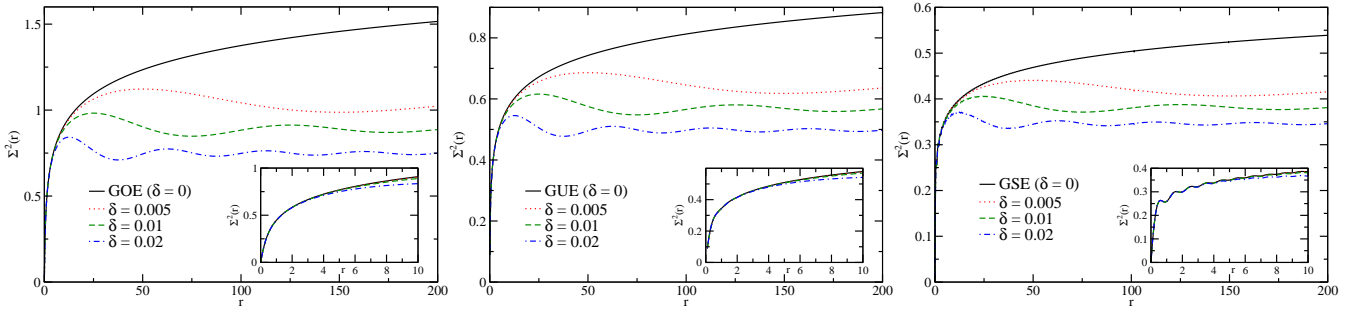

\subfloat{\includegraphics[width=0.33\textwidth, clip]{goe-model.eps}}
\subfloat{\includegraphics[width=0.33\textwidth, clip]{gue-model.eps}}
\subfloat{\includegraphics[width=0.33\textwidth, clip]{gse-model.eps}}
\caption{(Color online) Number variance for the model for interacting particle system discussed in Section \ref{egxe-sec-sigmasq-sat} for several values of $\delta$. Inset of each Figure shows the corresponding short range behavior. \label{egxe-fig-sigmasq-model}}
\end{figure*}

We also mention that the suppression of short wave frequencies should occur in the case of interacting embedded ensembles also \cite{embedded-2, rmp-note} and, therefore, we expect (\ref{egxe2-sigmasq-sat}) to apply again. In these cases, it is known from numerical simulations that the short-range fluctuations are identical to those of Gaussian ensembles \cite{embedded-3}. By generalizing (\ref{egxe-f2-model}), we can make a heuristic model for this case also. We consider the form factor
\begin{equation}
\mathcal F_2(k) = \left\{
\begin{array}{ll}
0, & 0 \leq |k| \leq \delta, \\
g(k), & \delta < |k| < \infty,
\end{array}
\right.
\end{equation}  
where $g(k) = 1-b_2(k)$ is the form factor of the Gaussian ensembles. Using (\ref{egxe2-sigmasq-sat}) we can predict that $\Sigma^2$ will become a constant for large $r$ with value given by,
\begin{equation}
\label{egxe-sigmasq-model-gxe}
\Sigma^2_{\mathrm{sat}}(r) =  2 \int_{\delta}^\infty \frac{g(k)}{2\pi^2 k^2} ~ \mathrm dk.
\end{equation}
We show $\Sigma^2$ in Fig.\ref{egxe-fig-sigmasq-model} using $g(k)$ of GOE, GUE and GSE, given ahead in (\ref{egxe-b2-goe}), (\ref{egxe-b2-gue}) and (\ref{egxe-b2-gse}) respectively. For $\delta > 0$, the number variance shows good agreement with the Gaussian ensemble results for $r \lesssim 10$ for the values of $\delta$ chosen. For larger values of $r$, saturation is observed and is consistent with (\ref{egxe-sigmasq-model-gxe}). We mention however that saturation in the interacting case has not yet been demonstrated in numerical simulations of embedded ensembles nor in the simulations of quantum chaotic systems. See however \cite{saturation-2, saturation-3} for saturation in the statistics of zeros of Riemann-$\zeta$ function.

In the earlier numerical and semiclassical calculations \cite{casati2, berry} for integrable and chaotic spectra of two degrees of freedom, saturation has been studied in terms of the rigidity statistics $\overline{\Delta_3}(r)$. This is related to $\Sigma^2(r)$ by \cite{apandey}
\begin{equation}
\overline{\Delta_3}(r) = \frac{2}{r^4} \int_0^r \left( r^3 - 2r^2s + s^3 \right) \Sigma^2(s)~ \mathrm ds,
\end{equation}
and therefore the corresponding saturation values are also related:
\begin{equation}
\overline{\Delta_3}_{\mathrm{sat}} = \frac{1}{2} \Sigma^2_{\mathrm{sat}}.
\end{equation}

\section{Embedded Gaussian Unitary Ensemble} \label{egxe2-sec-sigmasq-EGUE}
\begin{figure}[!ht]
\centering
\includegraphics[width=\linewidth, clip]{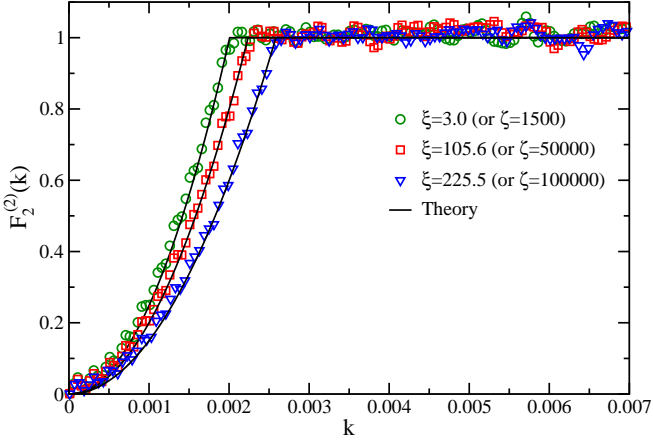}
\caption{(Color online) The two particle form factor for EGUE in different regions of spectra denoted by $\xi = 3.0, 105.6$ and $225.5$ respectively.\label{egxe2-fig-b2-EGUE}}
\end{figure}

We start with the simplest case, i.e., the EGUE. The form factor for the single particle spectrum is given by \cite{mlmehta},
\begin{equation}
\label{egxe-b2-gue}
b_2(k) = \left\{
\begin{array}{ll}
1-|k|, & |k| \leq 1,\\
0, & |k|\geq 1. 
\end{array}
\right.
\end{equation}
Substituting this in the expression for the two-level form factor given in (\ref{egxe-S2-2p}), we get,
\begin{equation}
\label{egxe-B2-EGUE}
F_2^{(2)}(k;\xi) = \left\{
\begin{array}{ll}
\chi, & |\chi| \leq 1,\\
1, &  |\chi| > 1,
\end{array}
\right.
\end{equation}
where $\chi = k R_1^{(2)}(\xi)$.
 We numerically calculate the form factor, $F_2^{(2)}(k;\xi)$ as discussed in Section \ref{egxe2-sec-mc}. The form factor at several locations in the two-particle spectra denoted by $\xi$ is shown in Fig.\ref{egxe2-fig-b2-EGUE}. The results are in excellent agreement with the analytical results given in (\ref{egxe-B2-EGUE}) and confirm the suppression of correlations for short frequencies ($|k| = O(N)$).

\begin{figure}[!ht]
\centering
\includegraphics[width=\linewidth, clip]{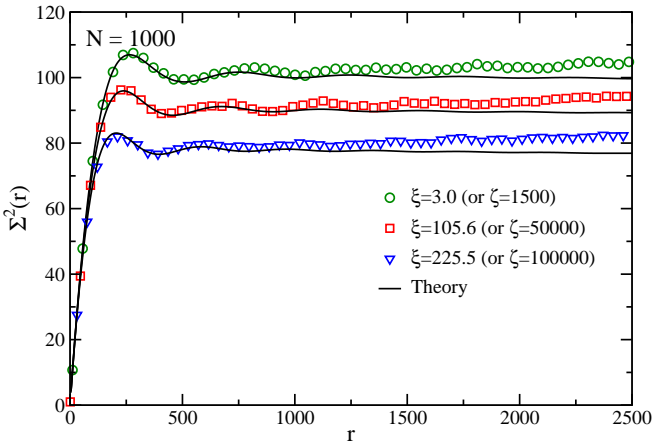}
\caption{\label{fig-gue-1000}(Color online) Number variance vs. $r$ for EGUE for three regions denoted by $\xi$. Solid lines are obtained from (\ref{egxe2-sigmasq-EGUE}).}
\end{figure}

For number variance, we substitute in (\ref{egxe2-sigmasq-B2}) the expression for the form factor given in (\ref{egxe-B2-EGUE}). See Appendix \ref{appendix-sigmasq-egue} for detailed calculation. We find,
\begin{align}
\label{egxe2-sigmasq-EGUE}
 \nonumber
 \Sigma^2(r;\xi) & = \frac{R_1^{(2)}(\xi)}{\pi^2}\bigg[ 2 + \pi^2 \Delta  - \cos(2\pi \Delta) - \frac{\sin(2\pi \Delta)}{2 \pi \Delta}  \\
& - 2\pi \Delta \Si(2\pi \Delta)  \bigg],
\end{align}
where $\Delta = r/R_1^{(2)}(\xi)$ and $\Si$ is the sine-integral function, defined as,
\begin{equation}
\label{egxe2-sinint}
\Si(x) = \int_0^x \frac{\sin(s)}{s} ~\mathrm d s.
\end{equation}
 Number variance, $\Sigma^2(r)$, is plotted against correlation length, $r$, in Fig.\ref{fig-gue-1000}. The agreement is very good for $r \approx N$. The small rise in $\Sigma^2$ for larger values of $r$ is due to finite size effect. We will see in Section \ref{egxe-sec-sigmasq-N5000} that the results become much better for $N = 5000$.

To find the number variance for small values of $r$, one can make Taylor expansion of (\ref{egxe2-sigmasq-EGUE}) around zero. The number variance for small values of $r$ can be written as,
\begin{equation}
\Sigma^2(r;\xi) = R_1^{(2)}(\xi) \left[\Delta - \frac{4}{3}\Delta^2 + \ldots\right].
\end{equation}
By using (\ref{egxe-zeta-xi-diff}), we get, 
\begin{equation}
\label{egxe2-sigmasq-EGUE-small}
\Sigma^2(r;\xi) = r - \frac{4}{3} \frac{1}{R_1^{(2)}(\xi)}r^2 + \ldots
\end{equation}
Since spectral density, $R_1^{(2)}(\xi)$, is $O(N)$, the number variance is linear for $r \ll N$, as verified in Fig.\ref{egxe-fig-sigmasq-smr-all}.

Number variance saturates at large correlation lengths, i.e., $r = O(N)$. We expand sine integral function given in (\ref{egxe2-sinint}) up to a few leading order terms,
\begin{align}
\label{egxe2-sinint-expand}
\nonumber
\Si(x)& = \frac{\pi}{2} - \int_x^\infty \frac{\sin(s)}{s} ~\mathrm ds \\
& = \frac{\pi}{2} - \frac{\cos(x)}{x} - \frac{\sin(x)}{x^2} + O(x^{-3}).
\end{align}
Substituting (\ref{egxe2-sinint-expand}) in (\ref{egxe2-sigmasq-EGUE}) and ignoring the lower order terms, we get the saturation value for $r \gtrsim N$,
\begin{equation}
\Sigma^2_{\mathrm{sat}}(r; \xi) = \frac{2}{\pi^2}R_1^{(2)}(\xi),
\end{equation}
which is same as derived from (\ref{egxe2-sigmasq-sat}).

\section{Embedded Gaussian Symplectic Ensemble} \label{egxe2-sec-sigmasq-EGSE}
\begin{figure}[!ht]
\centering
\includegraphics[width=\linewidth, clip]{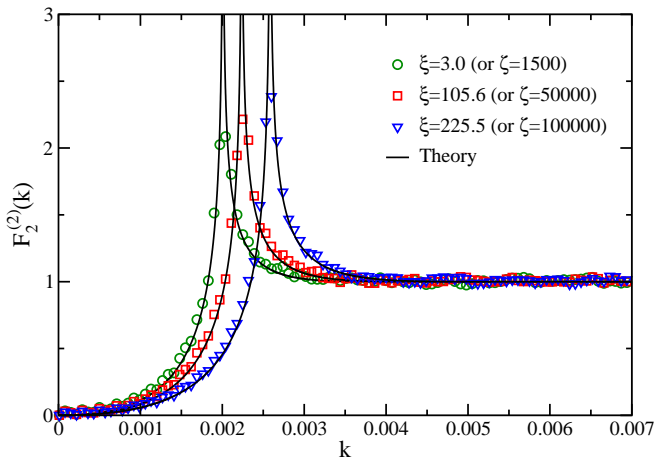}
\caption{(Color online) Two particle form factor for the EGSE for three values of $\xi$. \label{egxe2-fig-b2-EGSE}}
\end{figure}

For the Gaussian symplectic ensemble, the form factor, $b_2(k)$ is given by \cite{mlmehta},
\begin{equation}
\label{egxe-b2-gse}
b_2(k) = \left\{
\begin{array}{ll}
1-\frac{1}{2}|k| + \frac{1}{4}|k|\log(|(1-|k|)|), & |k| < 2,\\
0, & |k| > 2.
\end{array}
\right.
\end{equation}
By using (\ref{egxe-B2-2p}), the two particle form factor can be written as,
\begin{equation}
\label{egxe-B2-EGSE}
F_2^{(2)}(k;\xi) = \left\{
\begin{array}{ll}
\left[\frac{|\chi|}{2}  - \frac{|\chi|}{4} \log\left(\left|(1-|\chi|)\right|\right)\right]^2, & |\chi| < 2,\\
1, & |\chi| > 2,
\end{array}
\right.
\end{equation}
where $\chi = k R_1^{(2)}(\xi)$.
We calculate the form factor, $F_2^{(2)}(k;\xi)$ using (\ref{egxe2-S2}). The numerical results are displayed in Fig.\ref{egxe2-fig-b2-EGSE}. We observe good agreement with the analytical expressions given in (\ref{egxe-B2-EGSE}). Note that $F_2^{(2)}$ becomes singular for $|k|R_1^{(2)}(\xi) = 1$ as also does $1-b_2(k)$ at $|k| = 1$ for the one-particle case in (\ref{egxe-b2-gse}).

\begin{figure}[!ht]
\centering
\includegraphics[width=\linewidth, clip]{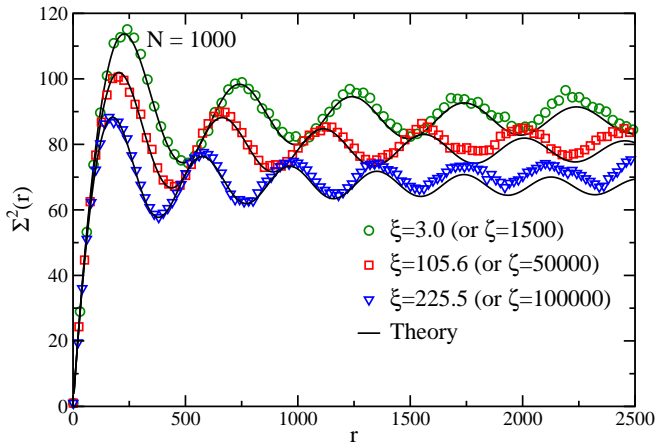}
\caption{(Color online) Variation of number variance with $r$ for EGSE. The number variance is calculated for various values of $\xi$. \label{egxe2-fig-sigmasq-EGSE}}
\end{figure}

We evaluate number variance by substituting the form factor for EGSE from (\ref{egxe-B2-EGSE}) in (\ref{egxe2-sigmasq-B2}). The calculation for evaluating number variance is given in Appendix \ref{appendix-sigmasq-egse}. We finally get the following expression for the number variance,
\begin{align}
\label{egxe2-sigmasq-EGSE}
\nonumber
\Sigma^2(r;\xi) & = \frac{R_1^{(2)}(\xi)}{8\pi^2}\left[ 14 + 8\pi^2\Delta - \frac{2\cos(2\pi\Delta)\Si(2\pi \Delta)}{\pi \Delta}      \right.\\
\nonumber &  - 16\pi \Delta \Si(4\pi \Delta)   - 4\cos(4\pi\Delta)   - \frac{\sin(4\pi \Delta)}{\pi \Delta}         \\
&\left. - 2\cos(2\pi\Delta)~ \pFq{2}{3}{\frac{1}{2},\frac{1}{2}}{\frac{3}{2},\frac{3}{2},\frac{3}{2}}{-\pi^2 \Delta^2}  \right].
\end{align}
Here $_2F_3$ represents the hypergeometric function defined as,
\begin{equation}
\label{hypergeometric-2f3}
\pFq{2}{3}{a_1, a_2}{b_1, b_2,b_3}{x} = \sum_{k=0}^\infty \frac{(a_1)_k (a_2)_k}{(b_1)_k (b_2)_k (b_3)_k} \frac{x^k}{k!},
\end{equation}
where $(a)_n$ is the Pochhammer symbol, given by
\begin{equation}
\label{pochhammer}
(a)_n = \frac{\Gamma(a+n)}{\Gamma(a)}.
\end{equation}
The number variance is shown in Fig.\ref{egxe2-fig-sigmasq-EGSE}. The numerical results follow the analytical expression very well. Again we consider the $N = 5000$ case in Section \ref{egxe-sec-sigmasq-N5000} for better agreement for large $r$ values.

Behavior at small correlation lengths is obtained by evaluating the Taylor expansion of the above expression around $r = 0$. The number variance for small $r$ can be written as,
\begin{equation}
\Sigma^2(r;\xi) = R_1^{(2)}(\xi)\left(\Delta - \frac{28}{27}\Delta^2 \ldots \right).
\end{equation}
Substituting (\ref{egxe-zeta-xi-diff}) in the Taylor expansion, we get, 
\begin{equation}
\Sigma^2(r;\xi) = r - \frac{1}{R_1^{(2)}(\xi)} \frac{28}{27}r^2 + \ldots
\end{equation}
Again since $R_1^{(2)}(\xi) = O(N)$, $\Sigma^2(r)$ is linear in $r$ for $r \ll N$. We have shown the linear behavior of number variance for EGSE at small values of $r$ in Fig.\ref{egxe-fig-sigmasq-smr-all}. The saturation value of $\Sigma^2$ at large correlation lengths can be obtained by evaluating the limiting values of the higher order terms. Using (\ref{egxe2-sinint-expand}) and ignoring the lower order terms, we get the saturation value, given by
\begin{equation}
\Sigma^2_\mathrm{sat}(r, \xi) = \frac{7}{4\pi^2} R_1^{(2)}(\xi)~\mbox{for}~r \approx O(N),
\end{equation}
again consistent with (\ref{egxe2-sigmasq-sat}). Note that the hypergeometric function, $_2F_3$, converges to zero for large values of $r$.

\section{Embedded Gaussian Orthogonal Ensemble} \label{egxe2-sec-sigmasq-EGOE}
\begin{figure}[!ht]
\centering
\includegraphics[width=\linewidth, clip]{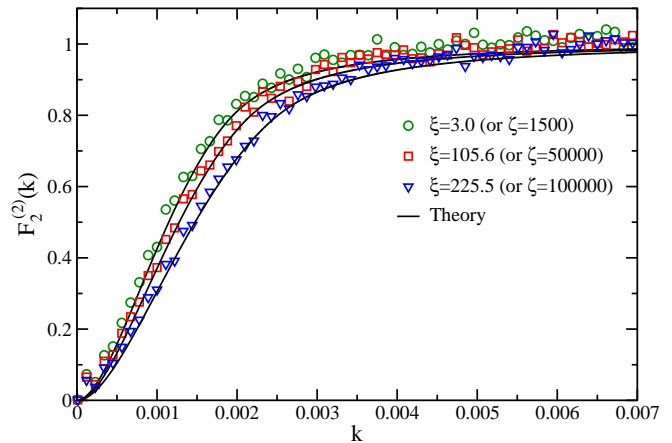}
\caption{(Color online) The two particle form factor for EGOE for three values of $\xi$. \label{egxe2-fig-b2-EGOE}}
\end{figure}

We consider the form factor for EGOE. The one particle form factor is given by,
\begin{equation}
\label{egxe-b2-goe}
b_2(k) = \left\{
\begin{array}{ll}
1 - 2|k| + |k| \log(1 + 2|k|), & |k| \leq 1,\\
-1 + |k|\log\left(\frac{2|k| + 1}{2|k| - 1}\right), & |k| \geq 1.
\end{array}
\right.
\end{equation}
Using (\ref{egxe-S2-2p}), we have for the two-particle system,
\begin{equation}
\label{egxe-B2-EGOE}
F_2^{(2)}(k;\xi) = \left\{
\begin{array}{ll}
\left[ 2|\chi| - |\chi| \log\left(1 + 2|\chi|\right)\right]^2, & |\chi| \leq 1,\\
\left[ 2 - |\chi|\log\left(\frac{2|\chi| + 1}{2|\chi| - 1}\right) \right]^2, & |\chi| > 1,
\end{array}
\right.
\end{equation}
where $\chi = k R_1^{(2)}(\xi)$.
We calculate the two-particle form factor, $F_2^{(2)}(k;\xi)$, using (\ref{egxe2-S2}). 
The form factor for EGOE at several values of $\xi$ is shown in Fig.\ref{egxe2-fig-b2-EGOE}. The numerical results are in good agreement with the analytical expression given in (\ref{egxe-B2-EGOE}).

\begin{figure}[!ht]
\centering
\includegraphics[width=\linewidth, clip]{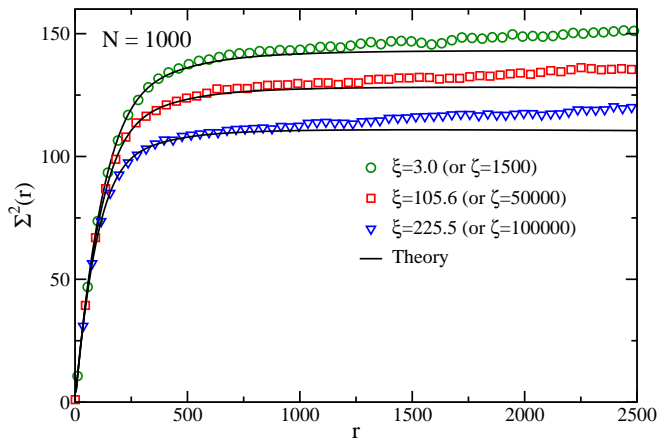}
\caption{(Color online) Number variance vs. $r$ for EGOE. The spectra are generated from the one particle spectra of size $N = 1000$. Number variance is evaluated for $\xi = 3.0, 105.6$ and $225.5$.}\label{egxe2-fig-sigmasq-EGOE}
\end{figure}

To evaluate the number variance, we substitute the two-particle form factor for EGOE from (\ref{egxe-B2-EGOE}) in (\ref{egxe2-sigmasq-B2}). The number variance can be written as,
\begin{widetext}
\begin{align}
\label{egxe2-sigmasq-EGOE}
\nonumber
\Sigma^2(r;\xi) & = \frac{R_1^{(2)}(\xi)}{2\pi^3\Delta}
\bigg[
\cos(\pi \Delta)[4\Si(\pi \Delta) + (\log 9 - 4)\Si(3\pi \Delta)]
-\sin(\pi \Delta)[4\Ci(\pi \Delta) + (\log 9 - 4)\Ci(3\pi \Delta)] \\
\nonumber
& - 2\pi \Delta \cos(2\pi \Delta)
 - (\log 3 - 2)^2 \sin(2\pi \Delta)
 + \pi \Delta [22 + 3\log3(\log3 - 6) - 4\pi \Delta \Si(2\pi \Delta)] - \pi \log3 \cos(\pi \Delta) \\ 
& + 2\pi^3\Delta^2 + 2\int_1^3 \int_\Delta^{\infty} \frac{\sin(\pi (kt-\Delta))}{kt}~\mathrm{d}k\mathrm{d}t
\bigg] + \sigma,
\end{align}
\end{widetext}
where
\begin{align}
\label{egxe2-sigmasq-EGOE-3}
\nonumber
\sigma & = 2 R_1^{(2)}(\xi) \int_1^\infty \Bigg[ 3 + \left[k \log\left(\frac{2k+1}{2k-1}\right) \right]^2 \\
& -4 k \log\left(\frac{2k+1}{2k-1}\right)\Bigg] 
 \left( \frac{\sin(\pi k \Delta)}{\pi k}\right)^2 ~ \mathrm dk.
\end{align}
See Appendix \ref{appendix-sigmasq-egoe} for the derivation of (\ref{egxe2-sigmasq-EGOE}).
We numerically obtain the number variance for EGOE at several values of $\xi$. We show in Fig.\ref{egxe2-fig-sigmasq-EGOE} the numerical results and its verification with the analytical expression given in (\ref{egxe2-sigmasq-EGOE}). Note that we calculate $\sigma$ numerically while calculating number variance from (\ref{egxe2-sigmasq-EGOE}). Again much better agreement is found for $N = 5000$ as shown in Section \ref{egxe-sec-sigmasq-N5000}.

We calculate the number variance for small correlation lengths in Fig.\ref{egxe-fig-sigmasq-smr-all}. We obtain linear behavior as in the case of EGUE and EGSE.

For number variance at large correlation lengths, we solve (\ref{egxe-B2-EGOE}, \ref{egxe2-sigmasq-sat}) directly for large values of $\Delta$. We find,
\begin{equation}
\Sigma^2_{\mathrm{sat}}(r; \xi) = \frac{1}{2\pi^2}\left[28 - 18\log(3) - 12 \mathrm{Li}_2(-2) - 2\pi^2\right]R_1^{(2)}(\xi)
\end{equation}
for $r\approx O(N)$. Here $\mathrm{Li}_2(x) = \sum_{k=1}^\infty x^k/k^2$ is the Dilogarithm function.

\section{Number variance for N = 5000} \label{egxe-sec-sigmasq-N5000}
\begin{figure}[!ht]
\includegraphics[width=\linewidth, clip]{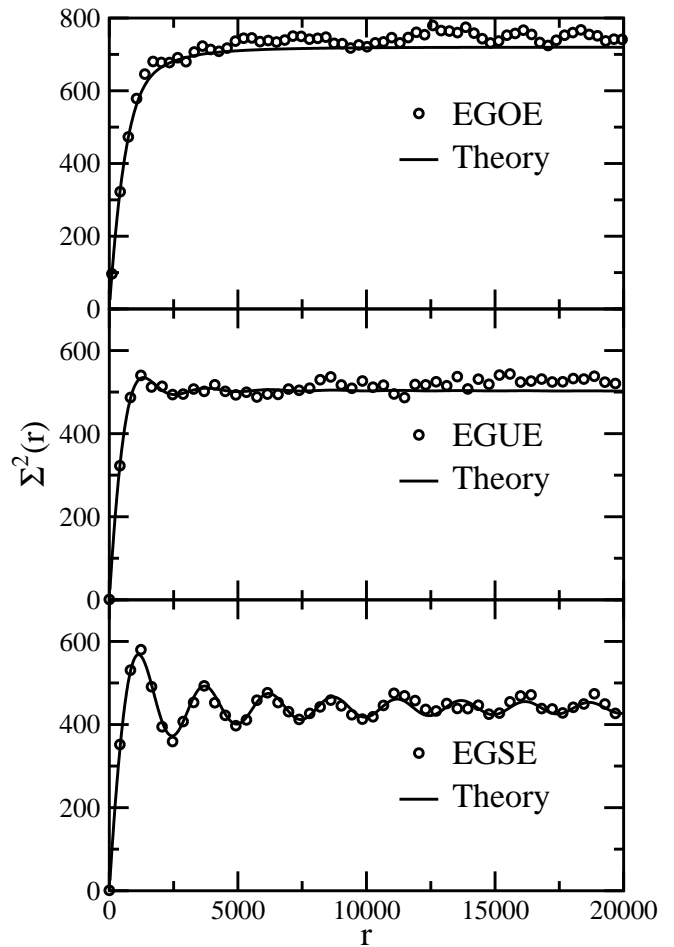}
\caption{Number variance for all three ensembles at $N = 5000$ and $\xi = 20.0$, which is equivalent to $\zeta = 50000$.\label{egxe-fig-sigmasq-N5000}}
\end{figure}

The number variance for all three ensembles, viz., EGOE, EGUE and EGSE at $N = 1000$ show small departures from the theoretical results for larger values of correlation lengths ($r \gtrsim N $). This is a finite-$N$ effect. To improve the agreement, we evaluate the number variance for $N = 5000$ with ensemble having $5000$ matrices. We find that the number variance follows the theoretical results for much larger correlation lengths ($r = O(4N)$) as shown in Fig.\ref{egxe-fig-sigmasq-N5000}.

\section{Saturation of Number Variance for Non-Uniform Density} \label{egxe2-sec-sigmasq-nonuniform}
\begin{figure}[!ht]
\subfloat[Spectral density \label{gue-conv}]{\includegraphics[width= \linewidth, clip]{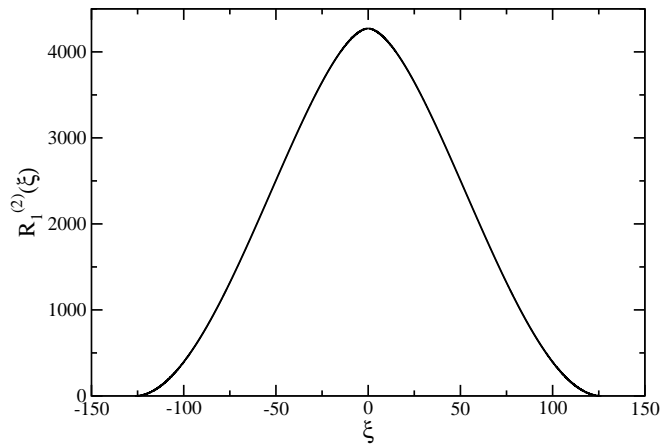}} \\
\subfloat[Number variance \label{egxe2-fig-sigmasq-EGUE-circle}]{\includegraphics[width= \linewidth, clip]{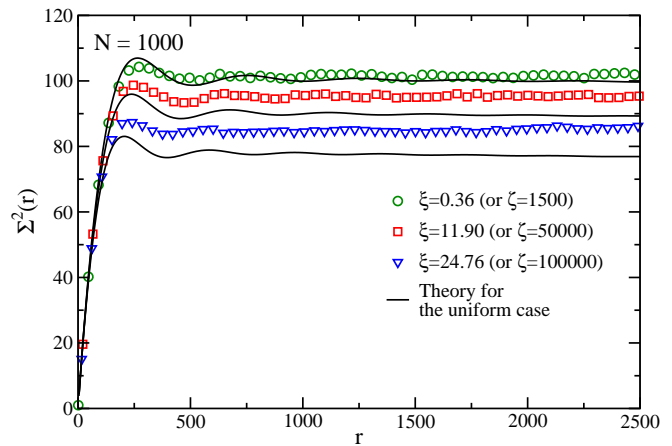}}
\caption{(Color online) Spectral density and number variance for two-particle spectrum constructed from one-particle spectra with semi-circular spectral density given in (\ref{gue-semicircle}) with $N = 1000$. Number variance is calculated in the same region on the unfolded scale where it was calculated in Fig. \ref{fig-gue-1000}. Data points denoted by circle, square and triangle represent the number variance around $\zeta = 1500, 50000$ and $100000$ respectively. Upper, middle and lower solid lines represent the number variance in the same region for the uniform case (i.e. for $R_1(x) = 1$).}
\end{figure}

We have considered above the number variance for the case when the one-particle spectral density is uniform. We now consider the number variance briefly for the case when the one-particle ensemble is GUE. Thus, the one-particle spectral density is semicircle and is given by
\begin{equation}
\label{gue-semicircle}
R_1(x) = \frac{1}{2\pi}\sqrt{4N - x^2}.
\end{equation}
We numerically obtain the two-particle density and also numerically unfold it to obtain number variance. The two-particle spectral density and number variance are shown in Fig.\ref{gue-conv} and Fig.\ref{egxe2-fig-sigmasq-EGUE-circle} respectively. Again we find saturation similar to (\ref{egxe2-sigmasq-sat}). Note that in both cases the data correspond to the same regions in the unfolded two-particle spectrum viz. $\zeta = 1500, 50000$ and $100000$ respectively for upper, middle and lower curves. We see that the saturation phenomenon is universal, but the actual numbers are different.

We emphasize that the $R_1^{(2)}$ used in all the expressions for $\Sigma^2, R_2^{(2)}, Y_2^{(2)}$ given in the earlier sections are valid only for the uniform case. The general theory that covers all types of one-particle density is yet to be worked out.

\section{Conclusion} \label{egxe-sec-summary}
We have studied the embedded ensembles for systems of two non-interacting particles. The one-particle spectra are modeled by (unfolded) Gaussian ensembles. We have derived correlation functions for such systems and studied the local fluctuation properties like nearest neighbor spacing distribution and the number variance. We find that for small correlation lengths two particle spectrum exhibits Poisson statistics which is a key behavior of integrable systems. For higher correlation lengths, the number variance approaches saturation values. We find that saturation (viz., linear to constant transition) occurs due to suppression of fluctuations of short-wave frequencies. Thus, spectra are rigid for such lengths. We study the number variance and their saturation values for all three types of embedded Gaussian ensembles, constructed using the spectrum of Gaussian ensembles viz., GOE, GUE and GSE, as the one-particle spectrum. The saturation phenomenon has been studied for chaotic systems also by semiclassical methods.  We present a heuristic model to illustrate the saturation phenomenon in the embedded ensembles of interacting particles. In these cases the spectrum displays a logarithmic to constant transition.

\acknowledgements
Ravi Prakash acknowledges CSIR, India for financial support.

\appendix

\section{Correlation function for the two-particle spectrum} \label{appendix-rn-2p}
The correlation functions can be written in terms of cluster functions using the definition given in (\ref{egxe-rn-tn}). One and two-level correlation functions are given in (\ref{egxe-r1-t1}) and (\ref{egxe-r2-t2}) respectively. Similarly, three and four-level correlation functions can be written as,
\begin{align}
\label{egxe-r3-t3}
\nonumber R_3(x_1,x_2,x_3) & = T_3(x_1,x_2,x_3) \\
\nonumber & -T_1(x_1)T_2(x_2,x_3) - T_1(x_2)T_2(x_1,x_3)\\
\nonumber &  - T_1(x_3)T_2(x_1,x_2) \\
& + T_1(x_1)T_1(x_2)T_1(x_3),
\end{align}
and
\begin{align}
\label{egxe-r4-t4}
\nonumber & R_4(x_1,x_2,x_3,x_4) = -T_4(x_1,x_2,x_3,x_4) \\
\nonumber & + T_1(x_1)T_3(x_2,x_3,x_4) + \ldots + \ldots \mbox{(4 terms)} \\
\nonumber & + T_2(x_1,x_2)T_2(x_3,x_4) + \ldots + \ldots \mbox{(3 terms)} \\
\nonumber & -T_1(x_1)T_1(x_2)T_2(x_3,x_4) - \ldots - \ldots \mbox{(6 terms)} \\
& + T_1(x_1)T_1(x_2)T_1(x_3)T_1(x_4).
\end{align}
Note that the $R_n$ are normalized to $N(N-1)\ldots (N-n+1)$. Similarly two-particle correlation function, $R_n^{(2)}$, are normalized to $d_2(d_2 - 1)\ldots (d_2 - n + 1)$.
 
 We will derive general expression for the spectral density and the two-level correlation function in terms of the correlation functions belonging to the single particle spectrum. The spectral density for the two-particle spectrum, $R_1^{(2)}(\xi)$, is given by
\begin{align}
\label{egxe-R1-2p-c2c1}
\nonumber
R_1^{(2)}(\xi) & = \int_{-\infty}^{\infty}   \int_{-\infty}^{\infty}  \left[C_2 R_2(x_1,x_2) + C_1 R_1(x_1)\delta(x_1 - x_2)\right] \\
& \times \delta\big(\xi - (x_1 + x_2)\big) ~ \mathrm dx_1 \mathrm{d}x_2.
\end{align}
The values of coefficients $C_1$ and $C_2$ in (\ref{egxe-R1-2p-c2c1}) are determined by the contribution of their corresponding correlation functions to the two particle spectrum. We know that two fermions can not occupy same energy level. Thus, we have, $C_1 = 0$. The normalization values on the left-hand side and right-hand side are $d_2 = N(N-1)/2$ and $C_2 N (N-1)$ respectively. We, therefore, have $C_2 = 1/2$. Substituting these values in (\ref{egxe-R1-2p-c2c1}), we get,
\begin{equation}
\label{egxe-R1-2p-r2}
R_1^{(2)}(\xi) = \frac{1}{2} \int_{-\infty}^{\infty}   \int_{-\infty}^{\infty}   R_2(x_1,x_2)~\delta\left(\xi - (x_1+x_2)\right) ~\mathrm dx_1 \mathrm dx_2.
\end{equation}
By substituting the relation between the correlation and cluster functions given in (\ref{egxe-r1-t1}) and (\ref{egxe-r2-t2}), we get the spectral density in terms of the cluster functions as given in (\ref{egxe-R1-2p-t2}) of the main text.

The two level correlation function, $R_2^{(2)}$, can be evaluated from the four level and lower-order correlation functions of the one particle spectrum. Thus $R_2^{(2)}$ can be expressed as a linear function of $R_4$ and $R_3$ as follows,
\begin{align}
\label{egxe-R2-2p-c4c3}
\nonumber
& R_2^{(2)}\left(\xi_1,\xi_2\right) =
 \int_{-\infty}^{\infty}  \int_{-\infty}^{\infty}  \int_{-\infty}^{\infty}  \int_{-\infty}^{\infty}
\big[C_4 R_4\left(x_1,x_2,x_3,x_4\right)  \\
\nonumber
&  + C_3 R_3\left(x_1,x_2,x_3\right) R_1(x_4)\delta(x_1 - x_4)\big] 
 \delta\big(\xi_1 - (x_1 + x_2)\big) \\
& \times \delta\big(\xi_2 - (x_3 + x_4)\big)~ \mathrm{d}x_1 \mathrm{d}x_2 \mathrm{d}x_3 \mathrm{d}x_4.
\end{align}
For Fermions, it can be assumed without any loss of generality that $x_1 > x_2$, $x_3 > x_4$ and $x_1 + x_2 \neq x_3 + x_4$. Under these conditions it can be easily verified that $R_2$ and $R_1$ can not contribute. Therefore, these are not considered in (\ref{egxe-R2-2p-c4c3}). The coefficients $C_3$ and $C_4$ are determined by their contributions to the two-particle correlation function. The normalization on left-hand side is equal to $d_2(d_2 - 1) = (N+1)N(N-1)(N-2)/4$. There are $N(N-1)(N-2)(N-3)/4$ permutations of $x_1,x_2,x_3,x_4$ such that all four $x$-variables are different. Therefore, these permutations must be equal to the normalization of $C_4 R_4$. The remaining $N(N-1)(N-2)$ permutations are such that two of the $x$-variables are same. So these permutations should be equal to the normalization of $C_3 R_3$. To keep the normalization equal to these values, we must have $C_4$ and $C_3$ equal to $1/4$ and $1$ respectively. Substituting these values in (\ref{egxe-R2-2p-c4c3}) and putting $R_1(x) = 1$, we get the expression for $R_2^{(2)}$ given in the main text by (\ref{egxe-R2-2p-r4}).

\section{Two-level correlation function for the two-particle spectrum} \label{appendix-r2-2p}
We write $R_2^{(2)}$ as the sum of two functions, $I_4$ and $I_3$, where $I_4$ is the integral over the $R_4$ term and $I_3$ is the integral over the $R_3$ term given in (\ref{egxe-R2-2p-r4}). Thus $R_2^{(2)}$ is given by
\begin{equation}
\label{egxe-R2-2p-I4I3}
R_2^{(2)} = \frac{1}{4}I_4 + I_3.
\end{equation}
 The order of integration of $T_3$ and $T_4$ terms is much lower than the remaining terms. We ignore these terms in further calculations.
 
 We first calculate $I_4$. We have
\begin{widetext}
\begin{align}
\label{egxe-I4-t2}
\nonumber I_4 &=
 \int_{-N/2}^{N/2}  \int_{-N/2}^{N/2}  \int_{-N/2}^{N/2}  \int_{-N/2}^{N/2}
   \Bigg[
    T_2(x_1,x_4) T_2(x_2,x_3) + T_2(x_1,x_3) T_2(x_2,x_4) 
 - \sum\limits_{i=1}^2 \sum\limits_{j=3}^4 T_2(x_i,x_j)
  \Bigg] \\
& \times     \delta(\xi_1 - (x_1 + x_2))  \delta(\xi_2 - (x_3 + x_4)) ~ \mathrm{d}x_1 \mathrm{d}x_2 \mathrm{d}x_3 \mathrm{d}x_4 
   + 4 R_1^{(2)}(\xi_1)~R_1^{(2)}(\xi_2).
\end{align}
\end{widetext}
Here we have used (\ref{egxe-R1-2p-t2}, \ref{egxe-r1-unfolded}) in (\ref{egxe-I4-t2}). The integrand in (\ref{egxe-I4-t2}) remains the same if we interchange $x_1$ with $x_2$ and $x_3$ with $x_4$. Therefore the first two terms give same values upon integration and the other four terms in the summation also yield same values. Thus, we get,
\begin{align}
\label{egxe-I4-t2-simple}
\nonumber I_4 & =
 \int_{-N/2}^{N/2}  \int_{-N/2}^{N/2}  \int_{-N/2}^{N/2}  \int_{-N/2}^{N/2}
 \big[
  2 T_2(x_1,x_3)T_2(x_2,x_4) \\
\nonumber
&   - 4 T_2(x_1,x_3) 
\big] 
 \delta(\xi_1 - (x_1 + x_2)) \\
\nonumber
& \times  \delta(\xi_2 - (x_3 + x_4)) ~\mathrm{d}x_1 \mathrm{d}x_2 \mathrm{d}x_3 \mathrm{d}x_4 \\
&  + 4R_1^{(2)}(\xi_1)~R_1^{(2)}(\xi_2).
\end{align}
For $\xi_1,\xi_2 \geq 0$ we find,
\begin{align}
\label{egxe-I4-t2-simple-center}
\nonumber I_4 & = \int_{-N/2 + \xi_2}^{N/2}  \bigg( \int_{-N/2 + \xi_1}^{N/2}
 \big[ 2 T_2(x_1,x_3)T_2(\xi_1 - x_1,\xi_2 - x_3) \\
&   - 4 T_2(x_1,x_3) 
   \big]~ \mathrm dx_1\bigg) \mathrm dx_3  + 4R_1^{(2)}(\xi_1)~R_1^{(2)}(\xi_2).
\end{align}
The one-particle cluster functions are translationally invariant for the unfolded spectra. Using (\ref{egxe-y2}), $I_4$ can be written as,
\begin{align}
\label{egxe-I4-y2}
\nonumber I_4 & = \int_{-N/2 + \xi_2}^{N/2}  \bigg( \int_{-N/2 + \xi_1}^{N/2}
 \big[
  2 Y_2(x_3 -x_1) Y_2\big(\Delta - (x_3 - x_1)\big) \\
&   - 4Y_2(x_1 - x_3)
   \big]
    ~\mathrm{d}x_1\bigg) \mathrm{d}x_3 
 + 4R_1^{(2)}(\xi_1)~R_1^{(2)}(\xi_2),
\end{align}
where $\Delta = \xi_2 -\xi_1$ as defined in (\ref{egxe-xi-def}). Away from the tail of the spectrum, using following change of variables,
\begin{align}
\label{egxe-I4-transformations}
s_1 = x_3 - x_1; ~ s_2 = x_3 + x_1
\end{align}
we can simplify (\ref{egxe-I4-y2}) considerably. We find,
\begin{align}
\label{egxe-I4-y2-simple}
\nonumber
I_4 & = 2(N - \xi)\int_{-\infty}^{\infty}Y_2(s_1)Y_2(\Delta - s_1)~\mathrm ds_1 - 4(N - \xi) \\
& + 4R_1^{(2)}(\xi_1)~R_1^{(2)}(\xi_2),
\end{align}
where $|\Delta| = O(1)$ and $\xi = (\xi_1 + \xi_2)/2$ as mentioned in (\ref{egxe-xi-def}).

Ignoring $T_3$ term, we can write $I_3$ as,
\begin{align}
\label{egxe-I3-t2}
\nonumber
 I_3 & = \int_{-N/2}^{N/2}  \int_{-N/2}^{N/2}  \int_{-N/2}^{N/2}  \int_{-N/2}^{N/2} 
   \big[
    1 - T_2(x_1,x_2) - T_2(x_1,x_3) \\
    \nonumber
    & - T_2(x_2,x_3)
     \big] \delta(x_1 - x_4)  
 \delta(\xi_1 - (x_1 + x_2)) \\
 \nonumber
& \times  \delta(\xi_2 - (x_3 + x_4)) ~ \mathrm dx_1 \mathrm dx_2 \mathrm dx_3 \mathrm dx_4.
\end{align}
simplifying as above, we find
\begin{equation}
\label{egxe-I3-y2}
I_3 = \int_{-N/2 + \xi_2}^{N/2} \big[1 - Y_2(2x-\xi_1) - Y_2(2x - \xi_2) - Y_2(\Delta)\big] ~\mathrm dx,
\end{equation}
and therefore for large $N$, we get,
\begin{equation}
\label{egxe-I3-y2-simple}
I_3 = (N-\xi + \frac{\Delta}{2})\left[1-Y_2(\Delta)\right].
\end{equation}
Substituting expressions for $I_4$ and $I_3$ in (\ref{egxe-R2-2p-I4I3}) and ignoring terms of $O(1)$, we get,
\begin{widetext}
\begin{equation}
\label{egxe-R2-2p-R1-2p}
R_2^{(2)}(\xi_1,\xi_2) - R_1^{(2)}(\xi_1)~R_1^{(2)}(\xi_2)= \frac{N - |\xi|}{2}\left[\int_{-\infty}^{\infty}Y_2(x)Y_2(\Delta - x)~\mathrm dx  - 2Y_2(\Delta)\right].
\end{equation}
\end{widetext}
This gives the cluster function for two-particle spectrum mentioned in (\ref{egxe-T2-2p}) of the main text. Note that the evaluation of (\ref{egxe-I4-t2-simple}) for $\xi_1, \xi_2 \leq 0$ also yields the same result. We have put $|\xi|$ in (\ref{egxe-R2-2p-R1-2p}) to include this case.

\section{Number variance for EGUE} \label{appendix-sigmasq-egue}
To evaluate the number variance, we substitute the expression for the form factor given in (\ref{egxe-B2-EGUE}) in (\ref{egxe2-sigmasq-B2}). We get
\begin{align}
\label{egxe-egue-eq1}
\nonumber
\Sigma^2(r;\xi) & = 
2\int_0^{1/R_1^{(2)}(\xi)} \left[ k R_1^{(2)}(\xi) \right]^2 \left( \frac{\sin(\pi k r)}{\pi k}\right)^2  ~ \mathrm d k \\
& + 2\int_{1/R_1^{(2)}(\xi)}^\infty \left( \frac{\sin(\pi k r)}{\pi k}\right)^2 ~ \mathrm d k.
\end{align}
Eq.(\ref{egxe-egue-eq1}) can be written as,
\begin{align}
\label{egxe2-sigmasq-egue-1}
\nonumber
\Sigma^2(r;\xi) & =
 2 R_1^{(2)}(\xi) \int_0^1 k^2 \left( \frac{\sin(\pi k \Delta)}{\pi k}\right)^2 ~ \mathrm dk \\
 & + 2R_1^{(2)}(\xi) \int_1^\infty \left( \frac{\sin(\pi k \Delta)}{\pi k}\right)^2 ~ \mathrm d k,
\end{align}
where
$r = R_1^{(2)}(\xi) \Delta$ as mentioned in (\ref{egxe-zeta-xi-diff}).
Note that,
\begin{equation}
\label{egxe2-sinint-integral}
\int_x^\infty \left(\frac{\sin(s)}{s}\right)^2 ~\mathrm ds = \frac{(\sin(x))^2}{x} + \frac{\pi}{2} - \int_0^{2x} \frac{\sin(s)}{s} ~\mathrm d s.
\end{equation}
 By substituting (\ref{egxe2-sinint-integral}) in (\ref{egxe2-sigmasq-egue-1}) we get the number variance given in (\ref{egxe2-sigmasq-EGUE}) of the main text.

\section{Number variance for EGSE} \label{appendix-sigmasq-egse}
We use the form factor for EGSE given in (\ref{egxe-B2-EGSE}) and substitute in (\ref{egxe2-sigmasq-B2}) to calculate the number variance. We get,
\begin{widetext}
\begin{align}
\nonumber
\Sigma^2(r;\xi) & =  2 R_1^{(2)}(\xi) \bigg[ \int_0^1\left [ \frac{|k|}{2}  - \frac{|k|}{4} \log\left(1-|k|\right)\right]^2 \left( \frac{\sin(\pi k \Delta)}{\pi k}\right)^2 ~\mathrm dk 
 +  \int_1^2 \left[\frac{|k|}{2}  - \frac{|k|}{4} \log\left(|k| - 1\right)\right]^2 \left( \frac{\sin(\pi k \Delta)}{\pi k}\right)^2 ~\mathrm dk\\
& +  \int_2^\infty \left( \frac{\sin(\pi k \Delta)}{\pi k}\right)^2~ \mathrm dk \bigg],
\end{align}
\end{widetext}
where $\Delta$ is related to $r$ as in (\ref{egxe-zeta-xi-diff}). We regroup and perform suitable transformations of variables to get,
\begin{align}
\label{egxe2-sigmasq-EGSE-2}
\nonumber
& \Sigma^2(r;\xi)  =
 \frac{R_1^{(2)}(\xi)}{2\pi^2} \int_0^2 \left[\sin(\pi k \Delta)\right]^2 ~\mathrm d k \\
\nonumber 
 & + \frac{2 R_1^{(2)} (\xi)}{\pi} \int_{2\pi \Delta}^\infty \left[ \frac{\sin(k)}{k}\right]^2 ~\mathrm d k \\
\nonumber
 & - \frac{R_1^{(2)}(\xi)}{2\pi^2} \int_0^1 \log(k) \left[ 1 - \cos(2\pi \Delta) \cos(2\pi \Delta k)\right] ~\mathrm d k \\
& + \frac{R_1^{(2)}(\xi)}{8 \pi^2} \int_0^1 [\log(k)]^2 \left[ 1 - \cos(2\pi \Delta)\cos(2\pi \Delta k)\right]~ \mathrm d k.
\end{align}
We mention the following standard integrals,
\begin{align}
\int_0^1 \log(t) \cos(a t) ~ \mathrm dt & = - \frac{\Si(a)}{a},\\
\int_0^1 [\log(t)]^2 \cos(a t) ~ \mathrm dt & = 2~ \pFq{2}{3}{\frac{1}{2},\frac{1}{2}}{\frac{3}{2},\frac{3}{2},\frac{3}{2}}{-\frac{a^2}{4}},\\
\end{align}
and
\begin{equation}
\int_0^1 \left[ \frac{1}{4}(\log(t))^2 - \log(t) \right] ~\mathrm dt = \frac{3}{2}.
\end{equation}
Here $_2F_3$ represents the hypergeometric function defined by (\ref{hypergeometric-2f3},\ref{pochhammer}).
We solve (\ref{egxe2-sigmasq-EGSE-2}) using these expressions to get the number variance given in (\ref{egxe2-sigmasq-EGSE}) of the main text.

\section{Number variance for EGOE} \label{appendix-sigmasq-egoe}
The number variance is evaluated by substituting the two-particle form factor for EGOE from (\ref{egxe-B2-EGOE}) in (\ref{egxe2-sigmasq-B2}). We get,
\begin{align}
\label{egxe2-sigmasq-EGOE-2}
\nonumber
\Sigma^2(r;\xi) & =  2 R_1^{(2)}(\xi)\int_0^1 \left[ 2|k| - |k| \log\left(1 + 2|k|\right)\right]^2 \\
\nonumber
& \times \left( \frac{\sin(\pi k \Delta)}{\pi k}\right)^2 ~\mathrm dk \\
\nonumber & + 2 R_1^{(2)}(\xi) \int_1^\infty   \left( \frac{\sin(\pi k \Delta)}{\pi k}\right)^2 ~ \mathrm dk.\\
& + \sigma,
\end{align}
where $\sigma$ is given in (\ref{egxe2-sigmasq-EGOE-3}).
 The integrals involved in (\ref{egxe2-sigmasq-EGOE-3}) are not easy. By numerical calculations, we find that the contribution from (\ref{egxe2-sigmasq-EGOE-3}) is of much lower order. Ignoring the contribution from $\sigma$, we get an error of about $6\%$. Therefore, we solve here (\ref{egxe2-sigmasq-EGOE-2}) without evaluating the $\sigma$ term. We take help of the following integrals in solving (\ref{egxe2-sigmasq-EGOE-2}),
\begin{equation}
\int \log(x) \sin (ax) ~\mathrm dx = \frac{\Ci(ax)}{ax} - \frac{\log(x) \cos(ax)}{a},
\end{equation}
\begin{equation}
\int \log(x) \cos (ax)~ \mathrm dx = \frac{\log(x) \sin(ax)}{a} - \frac{\Si(ax)}{ax}, 
\end{equation}
\begin{equation}
\int \left[ \log(x) \right]^2~ \mathrm dx = 2x - 2x \log(x) + x [\log(x)]^2,
\end{equation}
\begin{align}
\nonumber & \int [\log(x)]^2 \sin (ax) ~\mathrm dx = -[\log(x)]^2 \frac{\cos (ax)}{a} \\
& + \frac{2 \log(x)}{a} \Ci(a x) - \frac{2}{a}\int \frac{\Ci(ax)}{x} ~\mathrm dx,
\end{align}
\begin{align}
\nonumber & \int [\log(x)]^2 \cos (ax)~ \mathrm dx = [\log(x)]^2 \frac{\sin (ax)}{a} \\
&- \frac{2 \log(x)}{a} \Si(a x) + \frac{2}{a}\int \frac{\Si(ax)}{x} ~ \mathrm dx,
\end{align}
and
\begin{equation}
\int_{a/\pi}^\infty \frac{\sin(\pi s x - a)}{s x} ~\mathrm ds = \frac{\Ci(ax) \sin(a)}{x} - \frac{\Si(ax) \cos(a)}{x}.
\end{equation}
By substituting the above expressions in (\ref{egxe2-sigmasq-EGOE-2}), we get the number variance given in (\ref{egxe2-sigmasq-EGOE}) of the main text.

\bibliography{references.bib}

\end{document}